\documentclass[twocolumn,prl,aps,nofootinbib]{revtex4}
\usepackage{graphicx}

\begin{document}

\title{``Empty waves'', ``many worlds'', ``parallel lives'' \\and nonlocal decision at detection\footnote{\textbf{To appear in}: Is science compatible with free will? Exploring free will and consciousness in light of quantum physics and neuroscience, A. Suarez and
P. Adams (Eds.), Springer, New York, 2012, Chapter 5, Section 4.}}

\author{Antoine Suarez}
\address{Center for Quantum Philosophy, P.O.Box 304, 8044 Zurich,
Switzerland; suarez@leman.ch}

\date{April 8, 2012}

\begin{abstract}

I discuss an experiment demonstrating nonlocality and conservation of energy under the assumption that the decision of the outcome happens at detection. The experiment does not require Bell's inequalities and is loophole-free. I further argue that the local hidden variables assumed in Bell's theorem involve de Broglie's ``empty wave'', and therefore ``many worlds'' achieves to reconcile locality with the violation of Bell inequalities. Accordingly, the discussed experiment may be the first loophole-free demonstration of nonlocality.

\end{abstract}


\maketitle

Interference experiments can be considered the entry into the quantum world. Suppose for instance a Mach-Zehnder interferometer experiment in the case where only one photon at the time is impinging: In order to explain the interference one invokes the wave behavior, and when it comes down to the click of the detector one invokes the particle picture (\emph{wave-particle duality}). As it is well known, according to \emph{standard} quantum mechanics which detector clicks (the outcome) becomes determined at the detection. In fact, most physicists share this view also referred to as ``the collapse of the wavefunction at detection" by the Copenhagen (standard) interpretation. ``Outcome's decision at detection'' shall be the basic assumption in this paper.

\begin{figure}[t]
\includegraphics[width=80mm]{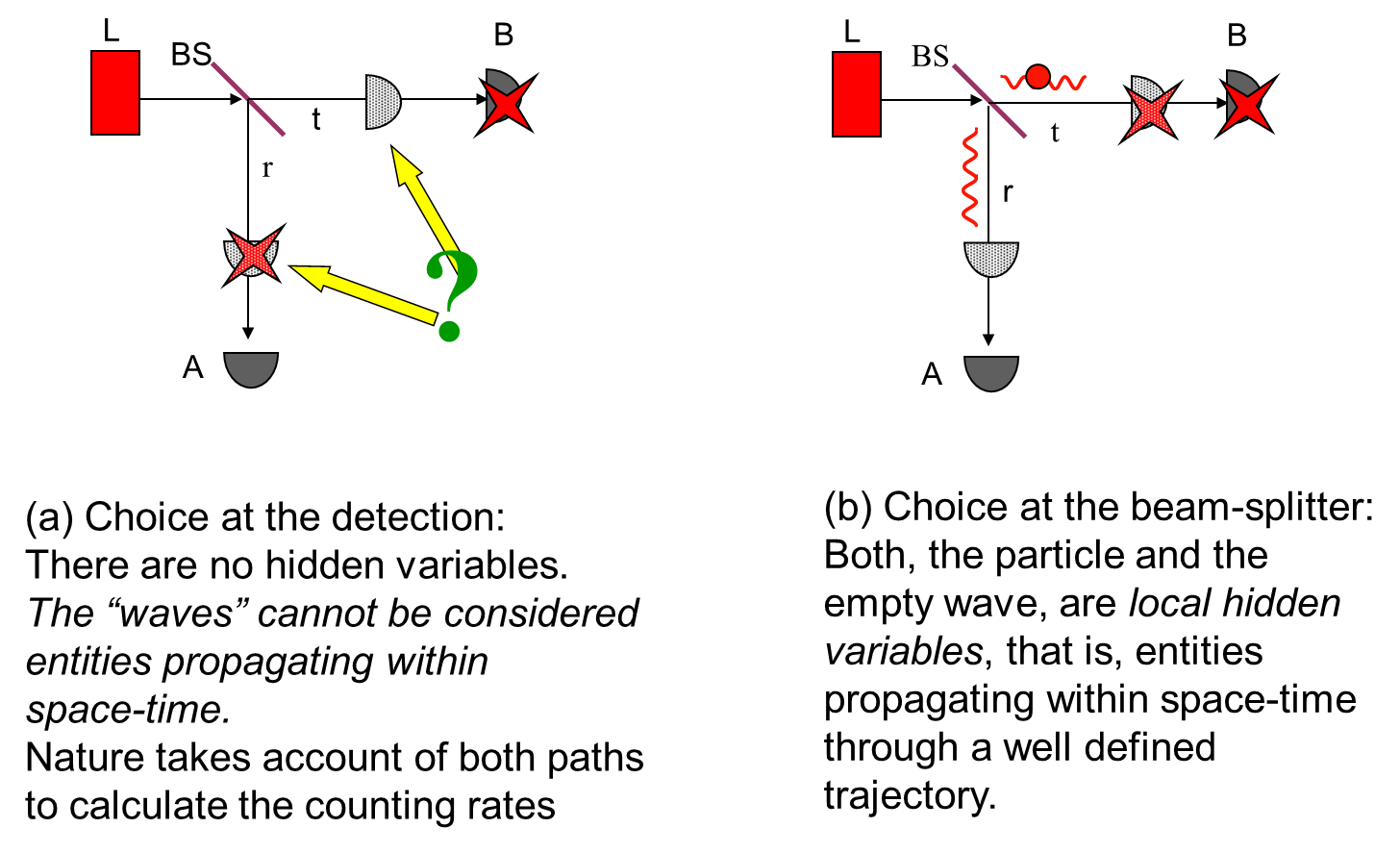}
\caption{ Gedanken-experiment: Suppose detector B (solid) fires in an experimental run. Would B have fired too, if the detectors (dashed) had been nearer to BS? a) According to ``choice at detection'' (collapse picture) there are no hidden variables or entities propagating in space-time, and one cannot conclude that B would have  fired. b) According to ``choice at the beam-splitter'' (de Broglie-Bohm's picture) there are hidden variables: The empty wave propagates in space-time, and detector B would also have  fired if the two detectors had been nearer to BS.}
\label{f11}
\end{figure}

Consider now the gedanken-experiment sketched in Figure \ref{f11}: A photon is impinging on a 50-50 beamsplitter BS with two output ports, one corresponding to the transmission path (\emph{t}) and the other to the reflection one (\emph{r}), and each port is monitored by a detector . According to the superposition principle we should say that the photon goes both paths, the transmitted and the reflected one. The outcome (which of the two detectors fires) will finally be decided at the detection.

An important implication of choice at detection is that the concept of trajectory doesn't make sense, as illustrated in Figure \ref{f11}(a): In case detector B fires in an experiment, one cannot conclude in a counterfactual experiment that B would also have fired had the two detectors been nearer to BS. However one could in principle assume that the collapse happens in a \emph{local} way, in the sense that an event cannot have effects elsewhere faster than light.

Suppose now that the two detectors are separated so far and at equal distances from the beamsplitter (equal optical path lenght) in a way that the two potential detection events are space-like separated. (This means that no communication can happen between the 2 detectors, informing one detector about the click or non-click of the other). Given this, from a \emph{local} point of view, we could reasonably assume that on each detector the photon has a $50\%$ chance to collapse, independently from the other detector. Then from time to time (in 1/4 of the cases) we will observe 2 clicks, and from time to time (in 1/4 of the cases) no click. This means an obvious violation of energy conservation in each single event, although energy would remain conserved in the average.

In this paper we discuss an experiment testing and ruling out this prediction, and this means in conclusion:
The assumption of the collapse of the wavefunction at detection implies that the principle of nonlocality emerges already in single-particle interference and rules the whole quantum physics. Additionally,``nonlocality at detection'' is necessary in order that energy remains conserved in each single quantum event.

\textbf{From ``empty waves'' to ``many worlds''}. Our experiment highlights that ``nonlocality'' was implicit in the Copenhagen interpretation of quantum mechanics, and this explains why the \emph{wave function collapse} raised Einstein's suspicion already in the Fifth Solvay Conference (1927) \cite{ja}, years before the celebrated EPR argument (1935). So, historically, nonlocality at detection appears before Bell's nonlocality (1964-1965) \cite{jb}.

Indeed the conclusion of ``nonlocal'' decision at detection fails if one assumes that the outcomes become determined at the beam-splitter BS. This was the basic assumption which led to the well known picture of ``the pilot or empty wave'' of Louis de Broglie \cite{jb,db}. According to this picture (Figure \ref{f11}(b)) there is always a particle traveling one path, and a pilot wave the alternative path. Although the pilot wave influences the particle and propagates in space-time, it is in principle ``empty'', that is inaccessible to observation or measurement. As John Bell himself emphasized, particle and empty wave build the set of local hidden variables that are ruled out by the violation of the Bell's inequality in entanglement experiments (see \cite{jb}, p. 128). Choice (of the outcome) at the beamsplitter is the natural consequence of assuming hidden variables, or more precisely, assuming hidden variables means nothing other than assuming ``empty waves''. By contrast, the assumption of the collapse at detection, presupposes that no hidden variables exist, and in this sense the local theory our experiment rules out is a \emph{local theory without hidden variables}.

Figure \ref{f11} illustrates the difference between the two views. It is interesting to note that both views assume entities (``waves'') that are \emph{inaccessible} to direct observation. However according to the standard view these entities cannot be said to exist within space-time, whereas the de Broglie's empty wave propagates within space-time through a well defined path very much the same way like its associated particle propagates through the alternative path: ``the wave is supposed to be just as 'real' and 'objective' as say the fields of classical Maxwell theory -although its action on the particles [...] is rather original'', John Bell says (\cite{jb}, p.128). We come later to this crucial difference again.

Bohm's picture added a ``nonlocal quantum potential'' to de Broglie's ``local pilot wave'', and could thereby account for the quantum correlations in 2-particle entanglement experiments, of course violating  the Bell's inequality. However with the ``many worlds" picture the story went further to provide a way for reconciling locality with the violation of Bell's inequality, simply by assuming that at each beam-splitter the world splits into two different worlds and each possible outcome becomes realized (see \cite{jb}, pp. 93-99, 192).``Many worlds'' addresses not only the ``measurement conundrum'' but also nonlocality.

As we will see later, if one assumes decision of the outcome at the beam-splitter, and therefore ``empty waves'', one cannot consistently reject ``many worlds'', and therefore the experimental violation of Bell's inequality, even without detection loophole, does not prove nonlocality. Alternatively, if one assumes decision of the outcome at detection, then one can escape ``many worlds'', and the natural and straightforward demonstration of nonlocality is provided by the experiment we present in the follow, which is a simplification of the experiment described in \cite{as10}.

\begin{figure}[t]
\includegraphics[width=80 mm]{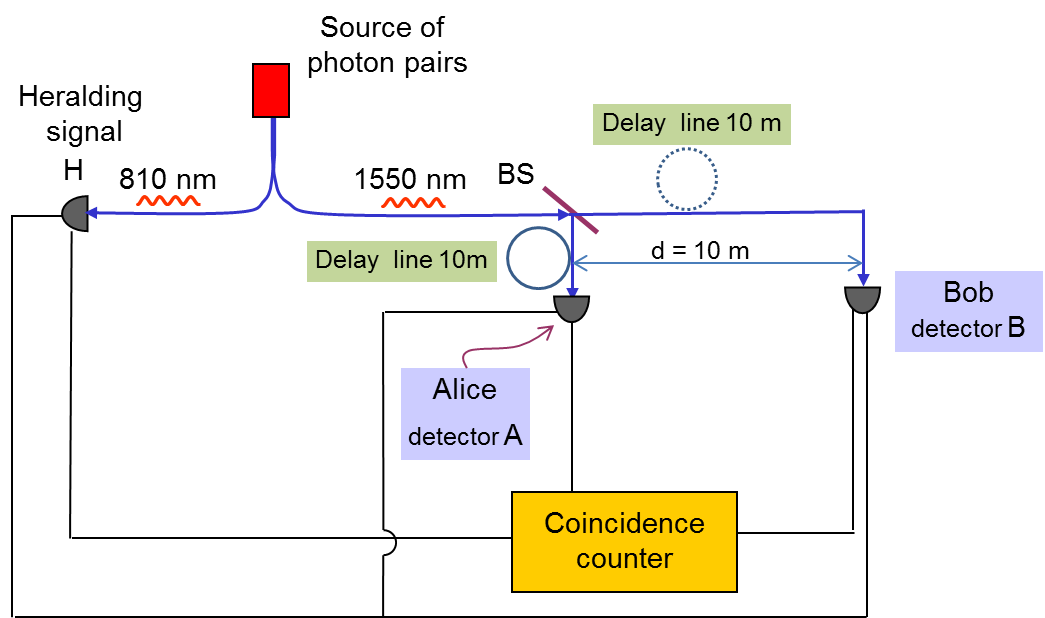}
\caption{Experiment demonstrating nonlocality at detection. A source generates pairs of photons at the wavelengths of 1550 nm and 810 nm. These pairs are split, and the 1550 nm photon is guided to a 50-50 beam-splitter BS and after leaving BS gets detected: Alice monitors one of the output paths of BS with detector A, and Bob the other output path with detector B. The 810 nm photon is sent to detector H, and used to herald the presence of the 1550 nm photon.}
\label{f2}
\end{figure}

\textbf{A loophole-free experiment}. If one takes for granted interferences effects \cite{as10}, then the principle of nonlocality at detection can be tested by the experiment proposed in Figure \ref{f2}. The technical aspects, the results, and some implications of the experiment have been presented in \cite{gszgs}. In the follow I focus on implications that have not been discussed so far.

We introduce the following notations:

$P(1,0)$: probability of getting a count in detector A and no count in B;

$P(0,1)$: probability of getting no count in A and one count in detector B;

$P(1,1)$: probability of getting one count in both detectors;

$P(0,0)$: probability of getting no count in any detector;

$P_A=P(1,0)+P(1,1)$: probability of getting a count in detector A;

$P_B=P(0,1)+P(1,1)$: probability of getting a count in detector B.

The photons are guided from BS to A and from BS to B by single mode optical fibers in two different configurations:

- \emph{Spacelike}: The delay line of $10m$ is set on Alice's path. The fibers from BS to A and from BS to B are equal in length, and both detectors A and B are separated from each other by a real distance of $d=10m$. Since the jitter time of the detectors is about $1ns$, the condition permitting signaling between A and B is given by $d\leq10^{-9}s\times3\cdot10^{8}\frac{m}{s}=0.3m$. That is: A and B are spacelike separated.

- \emph{Timelike}: The delay line of $10m$ is set on Bob's path. The glass fiber from BS to B is $20m$ larger than the fiber from BS to A, but both detectors A and B remain separated from each other by a real distance of $d=10m$. A and B can signal to each other during a time of $20m/(2\cdot10^{8}\frac{m}{s})=100ns$. Thus the condition permitting signaling between A and B is: $d\leq(100+1) 10^{-9}s\times3\cdot10^{8}\frac{m}{s}\approx30m$. That is: A and B are timelike separated.

One switches from the spacelike configuration to the timelike one simply by changing the delay line, without having to move detector B.

According to a \emph{local} theory, if the two detectors A and B are timelike separated, coordinated firing behavior is possible and the probabilities are given by:

\begin{footnotesize}
\begin{eqnarray}
&&P^{TL}(1,1)=P^{TL}(0,0)=0\nonumber\\
&&P^{TL}(1,0)=P^{TL}(0,1)=0.5\nonumber\\
&&P^{TL}_A=P^{TL}_B=0.5
\label{P}
\end{eqnarray}
\end{footnotesize}

\noindent where the  superscript \emph{TL} is used to denote the probabilities with timelike separation.

And if the two detectors A and B are spacelike separated, coordinated firing behavior is thwarted and the probabilities are given by:

\begin{footnotesize}
\begin{eqnarray}
&&P^{SL}(1,1)=P^{SL}(0,0)=P^{SL}(1,0)=P^{SL}(0,1)=0.25\nonumber\\
&&P^{SL}_A=P^{SL}_B=0.5
\label{P'}
\end{eqnarray}
\end{footnotesize}

\noindent where the superscript \emph{SL} is used to denote the probabilities with spacelike separation.

According to (\ref{P}) and (\ref{P'}) the \emph{local theory} yields the following predictions:

\begin{footnotesize}
\begin{eqnarray}
\frac{P^{SL}(1,1)}{P^{SL}_A \cdot P^{SL}_B}=1,\;\;\;\;\;\frac{P^{TL}(1,1)}{P^{TL}_A \cdot P^{TL}_B}=0
\label{PP'}
\end{eqnarray}
\end{footnotesize}

By contrast, the quantum mechanical probabilities remain invariant when one changes from timelike to spacelike configuration, and are identical to those predicted by the local theory for timelike separation. Accordingly the prediction of the \emph{quantum theory} for the spacelike separation is given by:

\begin{footnotesize}
\begin{eqnarray}
\frac{P^{SL}(1,1)}{P^{SL}_A \cdot P^{SL}_B}=0
\label{PQM}
\end{eqnarray}
\end{footnotesize}

We denote $R_{HA}$ the total number of coincident counts at detector H and detector A during the time of measurement, and $R_{H(A)}$ the total number of counts at detector H alone during the same measurement; $R_{HB}$ and $R_{H(B)}$ denote similar quantities for the measurement with H and B. $R_{HAB}$ denotes the number of triple coincident counts at the detectors H, A and B, and $R_{H(AB)}$  the total number of counts at detector H alone during the same measurement. All these quantities can directly be obtained by measurement.

These quantities are related to the probabilities in (\ref{PP'}) and (\ref{PQM}) by the equations:

\begin{footnotesize}
\begin{eqnarray}
P^{SL}_A = \frac{R_{HA}}{R_{H(A)}},\;\;\;P^{SL}_B = \frac{R_{HB}}{R_{H(B)}},\;\;\;P^{SL}(1,1)=\frac{R_{HAB}}{R_{H(AB)}}
\label{PR}
\end{eqnarray}
\end{footnotesize}

Similar relations hold for the probabilities and the counting rates measured with timelike separation.

Therefore, for \emph{spacelike} separation the local theory and standard quantum mechanics lead to two conflicting predictions (\ref{PP'}) and (\ref{PQM}), which can be tested experimentally. So, we have here a clear locality criterion allowing us to decide whether nature is nonlocal, provided the assumption of decision at detection.

According to quantum mechanics triple coincidence counts at the detectors H, A and B never happen if the source is ideal. However false triple coincidences will be observed due to probability of generating double pairs. Therefore one should expect to measure a very small $R_{HAB}$, both under timelike and spacelike separation. A source producing a large amount of false triple coincidences could be a \emph{loophole} characteristic for our experiment.

The experimental results of this experiment are presented in Table 1 of \cite{gszgs}. From these results one can conclude:

1. The value $P^{SL}(1,1)$ predicted by the local theory is three orders of magnitude ($10^3$) larger than the measured one.

2. The measured values $P^{SL}(1,1)$ (triple coincidences in H, A and B with spacelike configuration) and $P^{TL}(1,1)$ (triple coincidences in H, A and B with timelike configuration) are equal (within the statistical error).

3. The measured values $P^{SL}(1,1)$ and $P^{SL}_N(1,1)$ (triple coincidences happening because A and B detect photons coming from different pairs) are equal (within the statistical error).

This means that our experiment rules out the local theory, as concluded in \cite{gszgs}, but it additionally means that this experimental falsification \emph{is free from a loophole due to a bad source}, as it seems the only one that could impair the result.

\textbf{Implications}. According to the local theory, with spacelike separated detectors \emph{one single} photon produces two counts in $25\%$ of the events, and no count in other $25\%$, i.e., the energy is conserved on average, but not in each individual quantum process. The experimental falsification of this prediction means that conservation of energy in individual quantum processes is inseparably related to nonlocal decision at detection.

As said in the beginning, one could obviously object that it is possible to escape nonlocality at detection simply by assuming that the outcomes become determined at the beam-splitters. Then detections on one output port of a beam-splitter do not influence detections on the other output port. But such a local model necessarily involves local hidden variables of the type of de Broglie's ``empty pilot wave''. Thus, the ``local models'' addressed by the conventional Bell-type experiments are actually ``local empty wave models''. When implemented in entanglement experiments they involve correlated outcome decisions at two spacelike separated beam-splitters. Since ``local empty wave models''  fulfill the well known locality criteria of Bell's inequalities, they are refuted by the experimental violation of such inequalities \cite{jb}. Accordingly violation of Bell's inequalities does certainly imply that the decisions at two beam-splitters cannot be explained by ``local hidden variable models'' (like the empty wave), but this doesn't imply that such decisions cannot be explained by \emph{any} local model \cite{brr}.

Indeed the ``many worlds'' interpretation achieves to reconcile locality and violation of Bell inequalities. This state of affairs was already acknowledged by John Bell (see \cite{jb}, p. 192). Recently Gilles Brassard and Paul Raymond-Robichaud have elaborated ``many worlds'' to a theory of ``parallel lives'', which they also present as a ``belief'' shared by ``the \emph{strong} Faithful'' of the scientific community called ``Church of the Larger Hilbert Space". According to ``parallel lives'' only human observers and their apparatuses split. So for instance Alice with her apparatus lives inside a bubble and when she performs a measurement the bubble splits into two bubbles. Inside one bubble Alice sees one outcome; and inside the other bubble Alice's copy sees the alternative outcome:``From now on, the two bubbles are living parallel lives. They cannot interact between themselves in any way and will never meet again''. Then these authors argue that the theory of ``parallel lives'' reconciles violation of Bell's inequalities with ``a fully deterministic, strongly local and strongly realistic interpretation of quantum mechanics" \cite{brr}.

The theory of ``parallel lives'' has the great merit of highlighting the following issue: If one assumes that the decision of the outcomes happens at the beam-splitter, then one has to accept the ``empty wave'' and, at the end, one is led to local interpretations of quantum mechanics like ``many worlds'' and ``parallel lives'', which are compatible with the violation of Bell inequalities.

But why cannot ``many worlds'' and ``parallel lives'' be rejected, if one assumes ``empty waves''?

Because all these pictures share a \emph{rejection} of the following basic principle:

\emph{Principle A}: Any entity in space-time is in principle accessible to a human observer unless both (the entity and the observer) are spacelike separated. Or in other words, the only way to have \emph{inaccessibility} within space-time is through space-like separation.

I think \emph{Principle A} is the reasonable way of characterizing the contents of space-time, and it should be assumed by any sound scientific theory. In fact \emph{Principle A} is at the heart of the Copenhagen interpretation of quantum mechanics, and in particular of Bohr's view. And it seems to be shared by Einstein as well, who in fact disqualified the ``empty waves'' terming them as \emph{ghost fields}.  Therefore, for reasons of scientific coherence one should reject ``empty waves'', ``many worlds'' and ``parallel lives''. And in any case one cannot say that ``many worlds'' reconciles quantum mechanics and Einstein's local realism because in fact it is at odds with both.

Notice that in accord with \emph{Principle A} above (and with Bohr and Einstein) I too assume that all what is in space-time is accessible, even if (in opposition to Einstein) I don't accept that all what matters for physical reality is in space-time: In quantum experiments the observed results emerge always from outside space-time.

I would like to stress that it is not sufficient to assume ``free will'' in order to escape ``many worlds'' or ``parallel lives'' \cite{ng}: One has to reject ``empty waves''  as well, and therefore accept that decision of outcome happens at detection. My argument is as follows:

If we accept that nonlocal coordination of outcomes is in principle possible, by which particular reason should we then reject nonlocal decision at detection? Only because we assume that the outputs of devices are necessarily determined by some cause in the past light-cone. But then one must consequently also assume that the outputs of the experimenter's brain are predetermined, and therefore he has no free will. In other words, the three assumptions: free will, ``empty waves'' and nonlocality cannot hold together. And this means that for the sake of free will assuming ``empty waves'', and therefore ``many worlds'' and ``parallel lives'', is not better than assuming Gerard 't Hooft's \emph{superdeterminism}.

Strictly speaking, if one assumes decision of outcomes at the beam-splitter one can neither have experimenter's freedom nor prove nonlocality. By contrast, if one assumes decision at detection, then one can have both, experimenter's freedom and experimental demonstration of nonlocality \cite{gszgs}.

Are there reasons allowing us to prefer one assumption to the other? One such reason is obviously the wish for freedom. However I see another strong reason in favor of the standard view of decision at detection: \emph{Principle A}.

In summary, in accord with Einstein but in conflict with ``empty waves" and ``many worlds" I assume that the human observer can in principle access all what lies in spacetime (unless it is spacelike separated); and in conflict with both (Einstein and ``many worlds") I assume decision at detection coming from outside spacetime.

Now arises the question: Suppose we try to explain the quantum correlations in entanglement experiments by means of a \emph{local theory without hidden variables}. Suppose Alice sets her two detectors timelike separated, and Bob sets his detectors timelike separated as well (this condition excludes an experiment like the represented in Figure \ref{f2}). Suppose moreover that Alice and Bob are spacelike separated. Is it still possible \emph{under these conditions} to demonstrate nonlocality? The answer is: In principle YES, through the experimental violation of a Bell's inequality. However, as it is well known, for the time being such Bell-type experiments are not loophole-free because of the low detector's efficiency.

Thus, as regards to deciding whether nature is nonlocal, the violation of Bell's inequalities in entanglement does not prove anything beyond what our experiment with spacelike separated detector proves. The difference is that our experiment is loophole-free, and for the time being Bell-type experiments exhibit the detection loophole.

Against the presented demonstration of nonlocality one could argue as well that it is useless for quantum key distribution (QKD). However cryptography based on the protocol BB84 \cite{bb84} (for the moment the only marketable implementation of QKD) is in fact based on (single-particle) quantum interference or equivalent polarization effects. This means that the BB84 protocol works either because ``nonlocality at detection'' or the local ``empty wave''. It is in fact this sort of (unobservable and inaccessible) agency that ensures that quantum effects cannot be classically reproduced, and make it possible that BB84-QKD cannot in principle be classically eavesdropped. Consequently, as far as one considers the local ``empty wave'' a weird concept, the BB84 protocol is based on the nonlocality our experiment demonstrates.

For sure, the violation of Bell inequalities may allow us in future to certify genuine randomness in a more practical way than the nonlocality involved in interference experiments does \cite{pa}, but for the time being QKD based on the violation of Bell inequalities has not yet entered the market.

\textbf{Conclusion}. If one assumes that the decision of the outcome happens at detection, the experiment presented above is a clear demonstration of nonlocality (likely the first loophole-free one), and shows that this principle rules the whole quantum physics: it emerges already in interference phenomena involving only two detectors and not only when four ore more detectors are involved. The experiment demonstrates also that conservation of energy in each single event implies nonlocal coordination of detections: the most fundamental principle ruling the material visible world emerges from non-material invisible principles.\cite{as12}

If one rejects the view that the outcome is decided at detection, then one has to accept de Broglie's ``empty wave'' and at the end ``many worlds'', where the experimental violation of Bell's inequality even without detection loophole does not prove nonlocality.

Our experiment shows that Bell's nonlocality emerges from nonlocality at detection. This may be the reason why quantum mechanics is not more nonlocal, or more precisely why the Tsirelson bound characterizes the world we live in \cite{as10a}.

One may wonder why till now one has omitted to perform the experiment presented above, especially if one considers that Einstein used it as gedanken-experiment to argue against quantum mechanics as early as 1927. This ``omission'' is telling us perhaps something interesting about ``hidden assumptions'' on the part of physicists: While using words like ``collapse of the wave function at detection'', in fact they kept instinctively thinking in images like the ``empty wave''.

The experiment presented in this paper looks like an amazing ``nonlocal Columbus' egg''.

\emph{Acknowledgments}: I am thankful to Antonio Acin, Bruno Sanguinetti, Gilles Brassard, Nicolas Gisin, Thiago Guerreiro, Paul Raymond-Robichaud and Hugo Zbinden for many stimulating discussions.

\end{document}